\documentclass[11pt]{article}

\usepackage{courier}
\usepackage{hyperref}
\usepackage{graphicx,subfigure}
\usepackage{epstopdf}
\usepackage{tikz}
\usepackage{amsmath}
\usepackage{multirow}
\usepackage{amsfonts}
\usepackage{float}
\usepackage{amssymb}
\usepackage{graphicx}
\usepackage{flushend}
\newtheorem{theorem}{Theorem}

\newtheorem{proposition}{Proposition}

\newtheorem{definition}{Definition}
\newtheorem{remark}{Remark}
\newtheorem{example}{Example}
\makeatletter

\newcommand{\Rmnum}[1]{\expandafter\@slowromancap\romannumeral #1@}

\makeatother

\AtBeginDocument{%
  \let\oldref\ref%
  \def\ref{\oldref*}}
  
\begin{document}
\title{\LARGE Perceptual Control with Large Feature and Actuator Networks}
\author{John Baillieul}
\maketitle
\let\thefootnote\relax\footnotetext{\noindent\underbar{\hspace{0.8in}}\\
John Baillieul is with the Departments of Mechanical Engineering, Electrical and Computer Engineering, and the Division of Systems Engineering at Boston University, Boston, MA 02115. John Baillieul may be reached at {\tt johnb@bu.edu}. \newline Support from various sources including the Office of Naval Research grants N00014-10-1-0952 and N00014-17-1-2075 is gratefully acknowledged. }

\begin{abstract}
This paper discusses elements of a control theory of systems comprised of networks of simple agents that collectively achieve sensing and actuation goals despite having strictly limited capability when acting alone.  The goal is to understand {\em sensorimotor} feedback control in which streams of data come from large arrays of sensors (e.g. photo-receptors in the eye) and actuation requires coordination of large numbers of actuators (e.g. motor neurons).  The context for this work is set by consideration of a stylized problem of robot navigation that uses optical flow as sensed by two idealized and precise photoreceptors.  A robust steering law in this setting establishes a foundation for exploiting optical flow based on averaged noisy inputs from large numbers of imprecise sensing elements.  Taking inspiration from neurobiology, the challenges of actuator and sensor intermittency are discussed as are learning actuator coordination strategies.  It is shown that there are advantages in having large numbers of control inputs and outputs.  The results will be shown to make contact with ideas from control communication complexity and the standard parts problem.
\end{abstract}

 \section{Introduction}

The growing interest in the theory and design of networked control systems is stimulated in part by a realization that complex networks of simple sensors, actuators, and communication links exhibit features that have no counterparts in traditional linear and nonlinear feedback designs.  In particular, important capabilities can be realized by interconnections of large numbers of rudimentary sensors and simple actuators that have very limited abilities when acting alone, but whose aggregated actions can carry out prescribed tasks.  Taking inspiration from neurobiology, what follows presents work with simple models aimed at understanding how imprecise measurements from an adequately diverse set of sensors and control actions taken by groups of limited authority actuators can collectively achieve system objectives.  Due to space limitations, the emphasis will be limited to groups of control actions.

Within the broad area of networked control systems, a large body of work has been focused on the emergence of consensus and polarization among groups of intercommunicating agents (\cite{Hu,Altafini,Proskurnikov,Parsegov}).  In these models, agents influence each other and evolve toward terminal states that indicate {\em consensus} (all states being the same) or {\em polarization} (individuals tending toward any of two or more distinct terminal states).  In models of {\em neuromorphic control} considered below, agents themselves don't evolve, but rather they operate collectively in affinity groups that are associated with system-wide goals.  Ongoing research is exploring how self organizing behavior and Hebbian learning is achieved by reinforcement of different patterns of agent activations.  The control mechanism to be considered is weighted voting in which the weight assigned to any agent's input is a function of how effectively that input contributes to successful completion of a prescribed task.

The overarching motivation for the research is the need for foundational principles and trustworthy algorithms to support autonomous mobility.  A particular line of inquiry has been focused on the use of optical flow sensing to generate steering commands for both air and ground vehicles, \cite{Sebesta,Kong,Seebacher,Corvese}.  Leveraging the capabilities of feature detectors, descriptors, and matchers that are now available in the {\em OpenCV} corpus (\cite{Canclini}), the research is aimed at understanding how to focus attention so as to mitigate the uncertainties of visual perception.  The goal is to achieve implementations that reflect the robustness of the stylized model discussed in the next section.  
Preliminary work in \cite{Baillieul1} examined the challenges of robot navigation based on reaction to rapidly evolving patterns of features in the visual field.  What follows below is an extension of this work.

The paper is organized as follows.  Section 2 provides a foundational result on the use of optical flow to robustly navigate between two parallel corridor walls.  The model is an extreme idealization in that it makes use of two perfect photoreceptors that compute optical flow associated with whatever wall point their gaze falls on.  Perfect flow sensing is not possible in either the natural world or implementations using computer vision, and the practical hedge for flow signals degraded by noise, data dropouts, and perceptual aliasing is the averaged weighting of flows associated with large numbers of visual features.  This ``wisdom of crowds'' is the basis of the approach to {\em neuromorphic} control that is discussed in the remainder of the paper.  Section 3 shows that the energy optimal cost of steering a finite dimensional linear system is always reduced by adding more input channels.  Section 4 introduces the {\em standard parts} approach to control in which the goal is to realize system motions by activating groups of fixed motion primitives.  The preliminary results described suggest a possible approach to what might be called Hebbian learning for control designs.  Future research is discussed in the concluding Section 5.

\section{Idealized optical flow based on two and many pixels}

\begin{figure}[h]
\begin{center}
\includegraphics[scale=0.45]{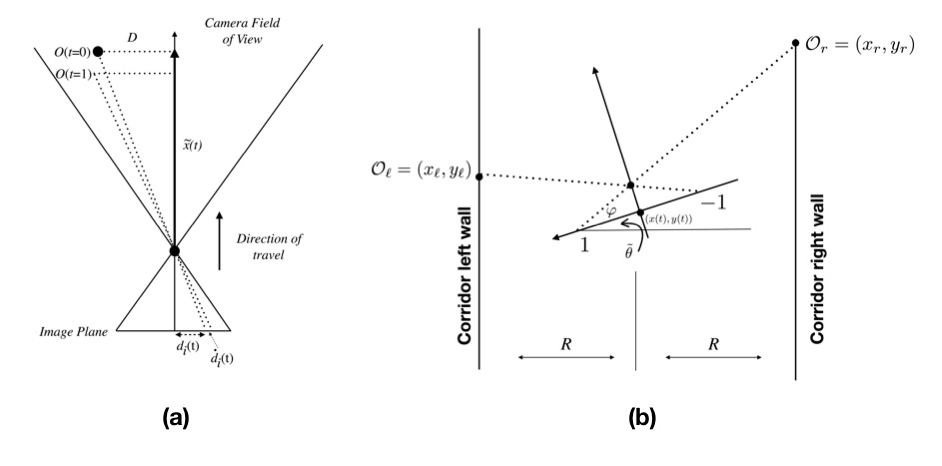}
\end{center}
\caption{Vehicle with kinematics (\ref{eq:jb:BasicVehicle}).  Here, the stylized image plane is compressed to one dimension and represented by a line segment on the vehicle $y$-axis. The vehicle $x$-axis is the direction of travel, $\tan\varphi=f$, the pinhole camera focal length, and $\theta=\tilde\theta+\frac{\pi}2$.  {\bf (a)} Time-to-transit, $\tau$, is available on the image plane as is evident from the geometry of similar triangles. {\bf (b)} The spatial features ${\cal O}_r$ and ${\cal O}_{\ell}$ are registered at $\pm 1$ respectively on the image plane ($y$-axis).}
\label{fig:RobotFigs}
\end{figure}

The optical flow based navigation discussed in \cite{Sebesta}, \cite{Kong} \cite{Seebacher}, and \cite{Corvese} exploits visual cues based on {\em time-to-transit} (sometimes referred to in the computer vision and visual psychology literature as {\em time-to-contact}, \cite{Horn}).  Time-to-transit (referred to in the literature as {\em tau}---and frequently denoted by the symbol $\tau$) is a perceived quantity that is believed to be computable in the visual cortex as animals move about their environments.  The basic idea is that when an object or feature is registered on the retina or image plane of a camera, the position of the image moves as a result of the animal's movement.  To explore the foundation of $\tau$-based navigation, we model a planar world in which  both environmental features and their images lie in a common plane.  A camera-fixed coordinate frame is chosen such that forward motion is always in the camera $x$-axis direction.  If we suppose that unobstructed forward movement is at a constant speed in a given direction, then as the camera moves through a field of environmental features, it  will at various points in time be directly abeam of each of the objects in its field of view.  The instant at which camera is directly abeam of an object or feature is referred to as the feature {\em transit time}.  At this point, as illustrated in Fig.\ \ref{fig:RobotFigs}{\bf (a)}, the object will lie instantaneously on the camera frame $y$-axis.  As seen in the figure. similar triangles may be constructed between the image plane and environmental scene through the focal point of the camera. Here, $D$ represents real world distance of an object from the axis of motion (camera frame $x$-axis) while $\tilde x(t)$ represents the distance to the point where the focal point will transit the feature (object).  Let $d_i(t)$ be the location of the object image on the camera image plane.  By comparing the ratios of corresponding sides in similar triangles, we see that

\begin{equation}
\frac{d(t)}{\tilde x(t)}=\frac{d_i(t)}{f}.
\label{eq:jb:similar}
\end{equation}
If we assume that $d(t)=D$ and speed $\dot x(t)=v$ are constant, then by differentiating the quantities in (\ref{eq:jb:similar}), we obtain
\begin{equation}
\dot d_i(t)\tilde x(t)-vd_i(t) = 0,
\end{equation}
and this may be rewritten as
\begin{equation}
\tau = \frac{\tilde x(t)}{v} = \frac{d_i(t)}{\dot d_i(t)}.
\label{eq:jb:tau}
\end{equation}
This shows that the {\em time-to-transit} $\tau$ is computable from the movement of pixels on the image plane or retina.  

Referring to Fig.~\ref{fig:RobotFigs}{\bf (b)}, we adopt a simple kinematic model of planar motion
\begin{equation}
\left(\begin{array}{c}
\dot x \\
\dot y \\
\dot\theta\end{array}\right) = \left(\begin{array}{c}
v\cos\theta \\
v\sin\theta \\
u\end{array}\right),
\label{eq:jb:BasicVehicle}
\end{equation}
where $v$ is the forward speed in the direction of the body-frame $x$-axis, and $u$ is the turning rate.  Following the derivation in \cite{Sebesta}, we find that the time-to-transit a feature located at $(x_r,y_r)$ in the world frame by a vehicle traveling at constant speed $v$ and having configuration $(x,y,\theta)$ (in world frame coordinates) is given by 
\begin{equation}
\tau(t) = \frac{\cos\theta(t)(x_r-x(t))+\sin\theta(t)(y_r-y(t))}{v}.
\label{eq:jb:tau1}
\end{equation}

The effectiveness of $\tau(t)$ as a steering signal can be easily established by considering an idealized visual model in which there are two perfect photoreceptors---one on each side of the center of focus.  In Fig.~\ref{fig:RobotFigs}{\bf (b)}, the photoreceptors are at $\pm 1$ along the camera frame $y$-axis. In the idealized model, it is assumed that instantaneous detections of perfectly clear features at both $d(t)=\pm 1$ and corresponding derivatives $\do d(t)$  are available to determine corresponding values of $\tau(t)$.  If the camera-carrying robot that is located between two long parallel walls and aligned roughly with the walls as will be made precise below, the left (right) photoreceptor (at corresponding body-frame point $(0,1)$ ($(0-1)$)) will register a unique position along the right (left) wall.  Assuming the vehicle moves forward at constant speed and that the feature point ${\cal O}_r=(x_r,y_r)$ is fixed in world frame coordinates and letting only $\theta$ vary, we find that $\tau=\tau(\theta)$ is maximized when the motion of the camera is aligned with its optical axis aimed directly at the feature point.  If two features ${\cal O}_{\ell}=(x_{\ell},y_{\ell})$ and ${\cal O}_r=(x_r,y_r)$ are located at exactly 
the same distance in the forward direction from the camera on opposite walls of a corridor structure as depicted in Fig.~\ref{fig:RobotFigs}{\bf (b)} ($y_{\ell}=y_r$), then we would find that $\tau_{\ell} = \tau_r$.  When this equality of left and right values of {\em tau} holds and the camera frame origin is nearer to one wall or the other, the orientation of the camera frame will have a heading that is directed away from the near wall.  Only when the vehicle is centered between the two walls will the vehicle heading be parallel to the walls.  Given these observations, the question we explore next is whether it is possible to steer the robot vehicle (\ref{eq:jb:BasicVehicle}) based on ego-centric comparisons of times to transit of feature points along opposite walls of the corridor.  This question is partially answered by the following.

\smallskip

\begin{theorem}
Consider a mobile camera moving along an infinitely long corridor with every point along both walls being a detectable feature that determines an accurate value of $\tau$.   As depicted in Fig.~\ref{fig:RobotFigs}{\bf (b)}, let $\tau_r=\tau({\cal O}_r)$ and $\tau_{\ell}=\tau({\cal O}_{\ell})$ be the respective times to transit the two feature points whose images appear at points equidistant on either side of the optical axis which defines the center of coordinates (focal point or {\em focus of expansion} (FOE)) in the image plane.  Then for any gain $k>0$ there is an open neighborhood $U$ of $(x,\theta)=(0,\frac{\pi}{2})$, $U\subset\{(x,\theta)\; :\; -R<x<R;\; \varphi<\theta<\pi-\varphi\}$ such that for all initial conditions $(x_0,y_0,\theta_0)$ with $(x_0,\theta_0)\in U$, the steering law
\begin{equation}
u(t)=k(\tau_{\ell}-\tau_r)
\label{eq:jb:tau-balance}
\end{equation}
asymptotically guides the vehicle with kinematics (\ref{eq:jb:BasicVehicle}) onto the center line between the corridor walls.
\label{th:jb:one}
\end{theorem}

\noindent{\em Proof:} \ 
There is no loss of generality in assuming that the image points at which transit times $\tau_r$ and $\tau_{\ell}$ are to be calculated lie at $\pm 1$ on the body frame $y$-axis (as depicted in Fig.\ \ref{fig:RobotFigs}{\bf (b)}).  It is also assumed that the corridor width, $2R$, is large enough to allow the vehicle to pass with some margin for deviation from a straight and centered path, and it is convenient to assume the vehicle travels at unit speed, $v=1$, in the forward direction.  (This means that the vehicle trajectory is parameterized by arc length.)  Letting $(f,0)$ be the body-frame coordinate of the pinhole lens in our idealized camera, we can locate the global frame coordinates of the wall features in terms of vehicle (camera-frame) coordinates $x,y,\theta$.  Specifically, elementary (but tedious) geometric arguments give the world coordinates corresponding to the feature images $(0,-1)$ and $(0,1)$ as:
\[
{\cal O}_{\ell}=\left(
\begin{array}{c}
 -R \\
 y+f \sin (\theta )+\frac{(R+x+f \cos (\theta )) (\cos (\theta )+f \sin (\theta ))}{\sin
   (\theta )-f \cos (\theta )} \\
\end{array}
\right)
\]
and
\[
{\cal O}_r=\left(
\begin{array}{c}
 R \\
 y+f \sin (\theta )+\frac{(R-x-f \cos (\theta )) (f \sin (\theta )-\cos (\theta ))}{f
   \cos (\theta )+\sin (\theta )} \\
\end{array}
\right)
\]
respectively.  Using (\ref{eq:jb:tau1}) with $v=1$, these coordinates yield corresponding geometric times to transit:
\[
\begin{array}{lc}
\tau_r&=\cos (\theta ) (R-x) + \sin (\theta ) \Big( f \sin (\theta )\\[0.07in]
&+\frac{(f \sin (\theta )-\cos (\theta )) (-f \cos
   (\theta )+R-x)}{f \cos (\theta )+\sin (\theta )} \Big)
   \end{array}
   \]
and
\[
\begin{array}{lc}
\tau_{\ell}&=\cos (\theta ) (-R-x)+\sin (\theta ) \Big(f \sin (\theta )
\\[0.07in]
&+\frac{(f \sin (\theta )+\cos (\theta )) (f \cos
   (\theta )+R+x)}{\sin (\theta )-f \cos (\theta )}\Big).\end{array}
   \]
From these values and some further algebra, we obtain
\begin{equation}
k(\tau_{\ell}-\tau_r) = -\frac{2 f k (f \cos (\theta ) (\sin (\theta )+R)+x \sin (\theta ))}{f^2 \cos ^2(\theta
   )-\sin ^2(\theta )}.
 \label{eq:jb:x-theta}
\end{equation}
To complete the proof, we note that the form of (\ref{eq:jb:x-theta}) allows us to isolate the subsystem
\begin{equation}
\left(\begin{array}{c}
\dot x \\
\dot\theta\end{array}\right) = \left(\begin{array}{c}
\cos\theta \\
k[\tau_{\ell}(x,\theta)-\tau_r(x,\theta)]\end{array}\right).
\label{eq:jb:reduced}
\end{equation}
It is easy to see that $(x,\theta)=(0,\pi/2)$ is a rest point for (\ref{eq:jb:reduced}), and indeed is the only rest point in the domain $\{(x,\theta)\; :\; -R<x<R;\; \varphi<\theta<\pi-\varphi\}$.   We can linearize about this point to get the first order approximation
\begin{equation}
\left(\begin{array}{c}
\dot{\delta x}\\
\dot{\delta\theta} \end{array}\right)=
\left(
\begin{array}{cc}
 0 & -1 \\
 2 f k & -2 \left(k f^2+k R f^2\right) \\
\end{array}\right)
\left(\begin{array}{c}
\delta x\\
\delta\theta \end{array}\right).
\label{eq:jb:linear1}
\end{equation}
The eigenvalues of the coefficient matrix are 
\begin{equation}
-f^2k(1+R)\pm\sqrt{fk[f^3k(1+R)^2-2]}.
\label{eq:jb:eigenvalues}
\end{equation}
These are always in the left half plane, proving the coefficient matrix is Hurvitz, and from this the theorem follows.
\begin{flushright}$\Box$\end{flushright}

\smallskip

\begin{remark} \rm
Letting the angle $\varphi$ be as depicted in Fig.\ \ref{fig:RobotFigs}{\bf (b)}, the condition that the initial orientation lies in the open interval $\varphi<\theta_0<\pi-\varphi$ is necessary.  Otherwise one of the walls will not register any features on one of the two image points (i.e., either $\pm 1$).  A margin for safety requires that $\varphi+\xi<\theta_0<\pi-\varphi-\xi$ for some $\xi>0$.  The values $\varphi$ and $\pi-\varphi$ are referred to as {\em critical heading angles}.
\end{remark}
\begin{remark} \rm
{\em The qualitative dynamics of two-pixel steering} (\ref{eq:jb:reduced}).  ({\em i}) By adjusting the gain $k$ in equation (\ref{eq:jb:tau-balance}), one has considerable control over the rate at which $(x,\theta)$ approaches $(0,\pi/2)$ corresponding to the vehicle aligning itself on the corridor center line.  ({\em ii}) While the eigenvalues (\ref{eq:jb:eigenvalues}) are always in the left half plane, in the range $0<k<2/(f^3(1+R)^2)$, they are complex numbers, implying that the linearized system (\ref{eq:jb:linear1}) oscillates.  Indeed, the vehicle undergoes small amplitude oscillatory motion settling onto the centerline trajectory.  ({\em iii}) As $k$ increases through the critical value $2/(f^3(1+R)^2)$, the eigenvalues become real, negative, and quickly diverge from one another in magnitude.  It is well known that when eigenvalues are negative with differing magnitudes that the phase portrait has trajectories aligned with the eigenaxis of the smaller magnitude eigenvalue.  This eigenaxis is a linear approximation to the tau-balance curve, $\tau_{\ell}(x,\theta)-\tau_r(x,\theta) = 0$ in the $(x$-$\theta)$-plane.
\end{remark}
\begin{remark} \rm
{\em The effect of delays and quantization}.  In laboratory implementations, the computations needed to implement the {\em sense--act} cycle in the steering law  (\ref{eq:jb:tau-balance}) are complex and may introduce latency.  Simulations of delays in the systems under discussion point to the need for soft gains $k$ in (\ref{eq:jb:reduced}).  Sampling also imposes limits as follows.
\end{remark}

\smallskip

\begin{theorem}
Consider the planar vehicle (\ref{eq:jb:BasicVehicle}) for which the steering law is of the sample-and-hold type:
\begin{equation}
u(t)=k[\tau_{\ell}(x(t_i),\theta(t_i))-\tau_r(x(t_i),\theta(t_i))], \ \ t_i\le t<t_{i+1},
\label{eq:jb:sampled}
\end{equation}
where the sampling instants $t_0<t_1<\dots$ are uniformly spaced with $t_{i+1}-t_i = h>0$.  Then for any sufficiently small sampling interval $h>0$, there is a range of values of the gain $0<k<k_{crit}$ such that the sampled control law (\ref{eq:jb:sampled}) asymptotically guides the vehicle with kinematics (\ref{eq:jb:BasicVehicle}) onto the center line between the corridor walls.
\label{thm:jb:sampled}
\end{theorem}

\noindent{\em Proof:} \ 
As in the previous theorem, we assume that the forward speed is constant ($v=1$).  We also assume a normalization of scales such that $f=1$.  It is again convenient to consider the angular coordinate $\phi=\theta-\pi/2$.  In terms of this, we have
\[
\dot\phi = k[\tau_{\ell}(x(t_i),\phi(t_i)+\pi/{2})-\tau_r(x(t_i),\phi(t_i)+{\pi}/{2})]
\]
on the interval $t_i\le t < t_{i+1}$.  Given the explicit formulas for $\tau_{\ell}$ and $\tau_r$, and given that the right hand side of the above differential equation is constant, we have the following discrete time evolution
\begin{equation}
\begin{array}{lcl}
\phi(t_{i+1})&=&\phi(t_i) \\[0.07in]
&&+ h k\frac{ 2 \sin \phi (t_1) (R+\cos \phi(t_i ))-2 x(t_i) \cos \phi(t_i )}{\sin ^2\phi (t_i)-\cos ^2\phi(t_i)
   }
   \end{array}
\end{equation}
In other words, the discrete time evolution of the heading $\phi$ is given by iterating the $x$-dependent mapping

\begin{equation}
g(\phi)=\phi+ h k\frac{ 2 \sin \phi  (R+\cos \phi)-2 x \cos \phi}{\sin ^2\phi-\cos ^2\phi.
   }
\label{eq:jb:iterate}
\end{equation}
Differentiating, we obtain
\begin{equation}\begin{array}{rr}
g^{\prime}(\phi) = 1 + \ \ \ \ \ \ \ \ \ \ \ \ \ \ \ \ \ \ \ \ \ \ \ \ \ \ \ &\\[0.04in]
\ \ \ \ \ \frac{2 h k (-2 -3 R \cos\phi +R \cos 3 \phi +3 x \sin \phi +x \sin3\phi
   )}{\cos ^2\phi -\sin ^2\phi }.&\end{array}
\label{eq:jb:quantized}
\end{equation}
The numerator is negative in the parameter range of interest, while the denominator is positive.  Hence, we can choose $k$ sufficiently small that $g$ is a contraction on $-\pi/4<\phi<\pi/4$ uniformly in $x$ in the range $-R<x<R$.  Thus the iterates of $\phi$ under the mapping (\ref{eq:jb:iterate}) converge to 0, and because $\dot x= -\sin\phi$, this proves the theorem. \ \ \ \ \ \ \ \hfill $\Box$

\smallskip

\begin{figure}[h]
\begin{center}
\includegraphics[scale=0.2]{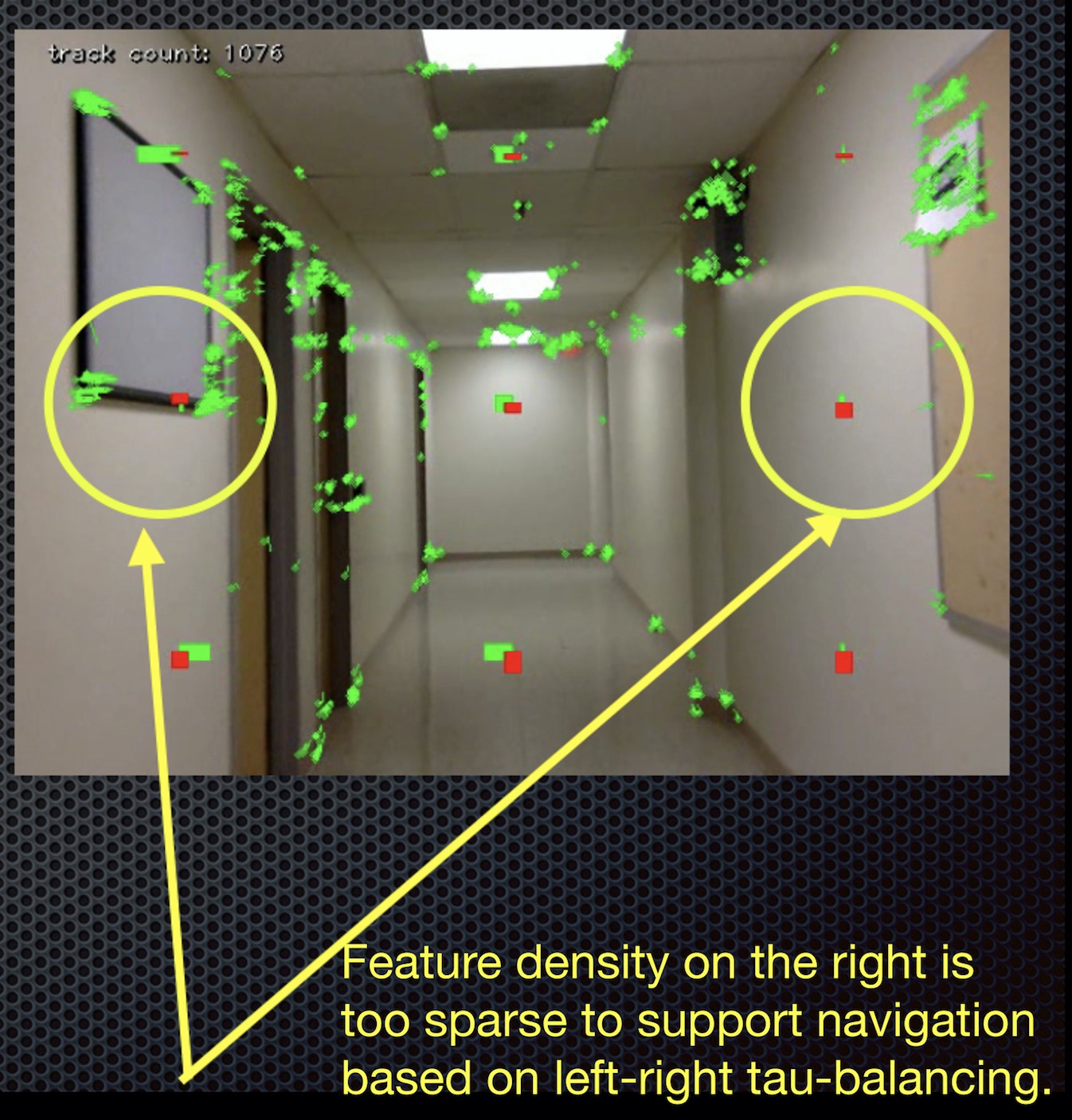}
\end{center}
\caption{Feature dropouts and sparse flow challenge visual navigation.}
\label{fig:SparseFlow}
\end{figure}

The model characterized by Theorem \ref{th:jb:one} is thus somewhat robust to both delays and input sampling.  It can also be shown to be quite robust with respect to gaze misalignment. (Think of one of the photo receptors being out of its nominal position at $y=\pm1$.). Apart from the simple temporal quantization implicit in sampling, spatial quantization of the perceived environment takes on paramount importance when vision based steering algorithms are designed.  In the corridor-following example above, if photoreceptors are located near the center of focus, the gaze will fall on wall points that may be at great distance from the current location.  There is considerable uncertainty in registering features that are far away due to the fact that tiny differences in pixel location near the center of focus correspond to very large differences in the spatial location of the features along the corridor.  Important open questions regarding optimal visual sampling of the field of view include how to balance input from very nearby features (which will have short time-to-transit and thus be relatively ephemeral), midrange features, and more distant features which may be important for the purpose of on-line replanning of motions.  Part of the challenge is selecting the portion of the field-of-view that requires the agent's primary attention.  In all cases, the environment as perceived by a camera is quantized into patches of pixels surrounding keypoints selected by feature detectors (e.g.\ SIFT, SURF, FAST, BRISK, and ORB from the OpenCV library).  In the approach described above, {\em tau}-values associated with the keypoints are computed, and the steering algorithms take weighted averages as input to the steering laws.  Whereas the idealized model of Theorem \ref{th:jb:one} gives precise input to center the vehicle between corridor walls, {\em tau}-values computed for the keypoints are subject to many types of errors including a typical rate of around 30\% failing to be matched from one frame to the next, \cite{Canclini}.  The reason that optical flow-based navigation works reasonably well is that if a large number of features are detected in each frame (around 2,000 in our implementations), these errors will average out.  As shown in Fig.\ \ref{fig:SparseFlow}, however, there may be portions of streaming video that have too few detectable features to provide a reliable steering signal. In such cases, algorithms that exploit alternative environmental cues are invoked.  Some of these are discussed in \cite{Kong,Corvese}, and others are the focus of ongoing research.  Continuing with analysis of highly idealized models, in the next section we consider the benefits of control designs for systems with relatively high dimensional input and output spaces.

\section{Control input and observed output---why more channels are better}

Because they are easy and revealing to analyze, we consider time-invariant (LTI) systems with $m$ inputs and $q$ outputs, whose evolution and output are given by
\begin{equation}
\begin{array}{l}
\dot x(t)=Ax(t) + Bu(t), \ \ \ x\in\mathbb{R}^n, \ \ u\in\mathbb{R}^m, \ {\rm and}\\[0.07in]
y(t)=Cx(t), \ \ \ \ \ \ \ \ \ \ \ \ \ \ y\in\mathbb{R}^q.\end{array}
\label{eq:jb:linear}
\end{equation}
As in \cite{Baillieul1}. we shall be interested in the evolution and output of (\ref{eq:jb:linear}) in which a portion of the input or output channels may or may not be available over any given subinterval of time.  Among cases of interest, channels may intermittently switch in or out of operation.  In all cases, we are explicitly assuming that $m > 1(\gg 1), q > 1 (\gg 1)$. Before addressing the problem of channel intermittency in Section 4, it will be shown that by increasing the number of input channels, the control energy is reduced.  To see this suppose the system (\ref{eq:jb:linear}) is controllable and consider the problem of finding the control input $u_0$ that steers the system from $x_0\in\mathbb{R}^n$ to $x_1\in\mathbb{R}^n$ so as to minimize
\[
\eta=\int_0^T\, \Vert u(t)\Vert^2\, dt.
\]
It's well known that $u_0$ may be given explicitly, and the optimal cost is 
\[
\eta_0 = (x_1-e^{AT}x_0)^TW_0(0,T)^{-1} (x_1-e^{AT}x_0)
\]
where $W_0$ is the controllability grammian:
\begin{equation}
W_0(0,T) = \int_0^T\, e^{A(T-s)}BB^T{e^{A(T-s)}}^T\, ds.
\label{eq:jb:grammian}
\end{equation}
(Here there is a slight abuse of notation in letting $T$ denote both a quatuty of time and the matrix transpose operator.  See e.g.\ \cite{Brockett}.)

\smallskip

\begin{theorem}
Suppose the system (\ref{eq:jb:linear}) is controllable with $A,B$  $n\times n$ and $n\times m$ matrices respectively.  Let $u_0(t)\in\mathbb{R}^m$ be the optimal control steering the system from $x_0\in\mathbb{R}^n$ to $x_1\in\mathbb{R}^n$ having minimum cost $\eta_0$ as above.
Let $\bar b\in\mathbb{R}^n$ and consider the augmented $n\times (m+1)$ matrix $\hat B=(B\vdots \bar b)$.  The $(m+1)$-dimensional control input $\hat u_0(t)$ that steers
\begin{equation}
\dot x(t)=Ax(t) + \hat B \hat u(t)
\label{eq:jb:augmented}
\end{equation}
from $x_0\in\mathbb{R}^n$ to $x_1\in\mathbb{R}^n$ so as to minimize
\[
\hat\eta=\int_0^T\, \Vert \hat u(t)\Vert^2\, dt
\]
has optimal cost $\eta_1\le\eta_0$.
\label{thm:jb:augmented}
\end{theorem}

\smallskip

\noindent{\em Proof:} \ 
It is easy to see that the augmented controllability grammian
\[
\begin{array}{rcl}
\hat W(0,T) & = & \displaystyle\int_0^T\, e^{A(T-s)}\hat B\hat B^T{e^{A(T-s)}}^T\, ds\\[0.2in]
& = & W_0(0,T) + W_p(0,T),
\end{array}
\]
where 
\[
W_p(0,T) =  \int_0^T\, e^{A(T-s)}\bar b\bar b^T{e^{A(T-s)}}^T\, ds.
\]
To simplify notation, denote these matrices by $W_0$ and $W_p$ respectively, and let $z=x_1-e^{AT}x_0$.  Then the optimal cost of steering the augmented system is $\eta_1 = z^T[W_0+W_p]^{-1}z$.

Since (\ref{eq:jb:linear}) is assumed to be controllable, $W_0$ has a positive definite square root; call it $S$.  Then write $\eta_1$ as
\[
\begin{array}{ll}
z^T[W_0+W_p]^{-1}z &
=z^T[S(I+S^{-1}W_pS^{-1})S]^{-1}z\\[0.07in]
&=z^TS^{-1}(I+M)^{-1}S^{-1}z,
\end{array}
\]
where $M=S^{-1}W_pS^{-1}$.  Because $M$ is positive semidefinite and symmetric, there is a proper orthogonal matrix $U$ and diagonal $\Delta$ with nonnegative entries such that
\[
\begin{array}{ll}
\eta_1 &= z^TS^{-1}(I+U^T\Delta U)^{-1}S^{-1}z \\[0.07in]
&=z^TS^{-1}[U^T(I+\Delta)U]^{-1}S^{-1}z\\[0.07in]
&=z^TS^{-1}U^T(I+\Delta)^{-1}US^{-1}z = y^T(I+\Delta)^{-1}y,
\end{array}
\]
where $y=US^{-1}z$.  Clearly $ y^T(I+\Delta)^{-1}y \le y^Ty$, and noting that $y^Ty=z^TS^{-1}U^TUS^{-1}z = z^TW_0^{-1}z$, this proves the theorem.
\begin{flushright}$\Box$\end{flushright}

\smallskip

\begin{remark} \rm
We assume that (\ref{eq:jb:linear}) is controllable so that $W_0$ is positive definite.  $W_p$ need not be p.d., however, but if it is, then the inequality of costs is strict: $\eta_1<\eta_0$.
\end{remark}

\begin{example} \rm
A simple illustration of Theorem \ref{thm:jb:augmented} is the planar special case of (\ref{eq:jb:linear}) where $A=\Big(\begin{array}{cc}
0 & 1\\ 0 & 0
\end{array}\Big)$ 
and $B=\Big(\begin{array}{c} 0 \\ 1 \end{array}\Big)$, $\Big(\begin{array}{cc} 0 &1 \\ 1 & 0\end{array}\Big)$, or $\Big(\begin{array}{ccc}
0 & 1 & 1\\
1& 0 & 1\end{array}\Big)$, representing the cases of one, two, or three input channels.  The system is ``optimally'' steered from the origin $(0,0)$ to points on the unit circle $(\cos\phi,\sin\phi)$ so as to minimize the $L_2$-norm on the control input.
\begin{figure}[h]
\begin{center}
\includegraphics[scale=0.43]{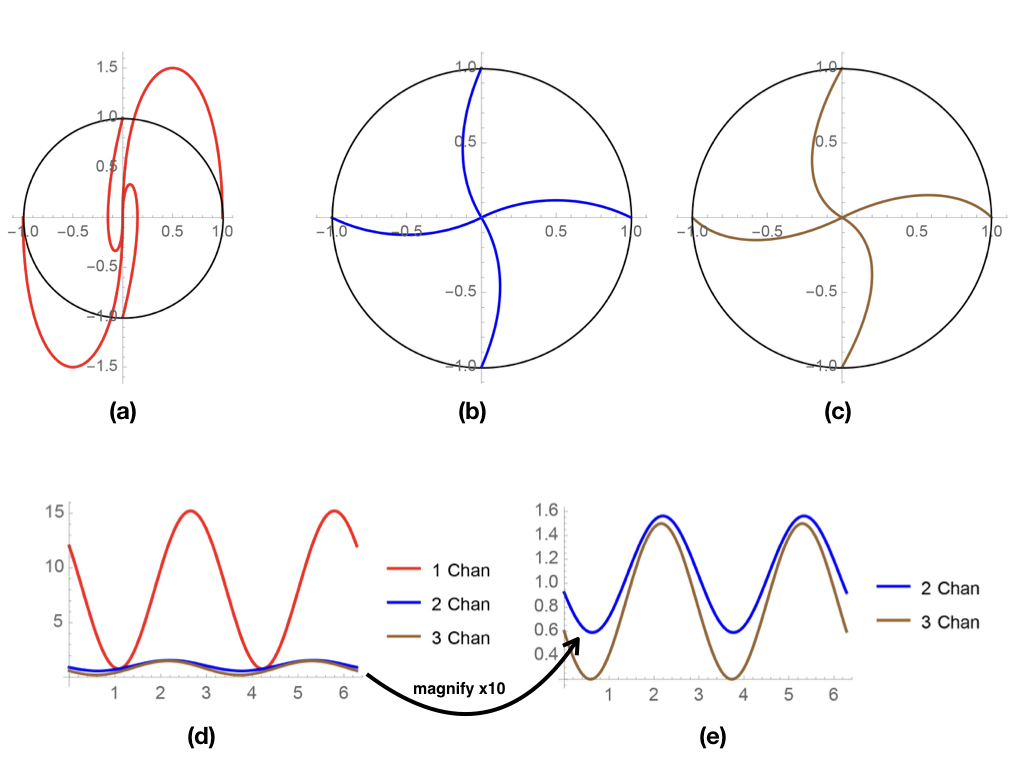}
\end{center}
\caption{The special case of (\ref{eq:jb:linear}) for a planar system with one, two and three input channels.
{\bf (a),(b)} and {\bf (c)} are trajectory plots of the one, two, and three channel systems corresponding to goal points at $(1,0),(0,1),(-1,0)$, and $(0,-1)$. {\bf (d)} and {\bf (e)} plot the costs of reaching points as a function of angular coordinate $\phi$.}
\label{fig:Channels}
\end{figure}
\end{example}

Fig.\ \ref{fig:Channels} illustrates optimal steering of (\ref{eq:jb:linear}) with these coefficient matrices to four points on the unit circle.  Trajectories for the one, two, and three channel cases are depicted in panels {\bf (a),(b),(c)} respectively.  Fig.\ \ref{fig:Channels}(a) is reduced in scale, and one sees a circuitous motion to the goal point.  Whereas the system is controllable using only the first channel of $\Big(\begin{array}{cc} 0 &1 \\ 1 & 0\end{array}\Big)$, it is not controllable using only the second channel.  Nevertheless, adding the second channel significantly changes the motion and dramatically lowers the energy cost $\eta$.  As Theorem \ref{thm:jb:augmented}
requires, the cost is further reduced by adding the third channel (panel {\bf (e)}), but the cost improvement is much less dramatic. 

\section{The Case of Large Numbers of Standard Inputs}

Having noted the control energy advantages of having large numbers of input channels, it is of interest in the context of neuroinspired models to consider the robustness inherent in having redundant channels as well as the possibility of designs in which groups of control primitives are chosen from catalogues and aggregated on the fly to achieve desired ends.  Biological motor control in higher animals is governed by networks of interconnected neurons, and drawing inspiration from the neural paradigm as well as from our recent work on {\em standard parts} control and control communication complexity (\cite{Wong1,Wong2}), we consider linear systems (\ref{eq:jb:linear}) in which control is achieved by means of selecting standard inputs from a catalogue of control functions that is large enough to ensure that the system can carry out a prescribed set of tasks.  Using a class of nonlinear models, this general approach was carried out with catalogues of sinusoids in \cite{Wong1} and Fourier series in \cite{Wong2}.  Here we briefly introduce dictionaries of {\em set-point stabilized} control primitives.  

Each control input will be of the form $u_j(t)=v_j+k_{j1}x_1 + \dots + k_{jn}x_n$ where the gains $k_{ji}$ are chosen to make the matrix $A+BK$ Hurwitz, and the $v_j's$ are then chosen to make the desired goal point $x_g\in\mathbb{R}^n$ an equilibrium of (\ref{eq:jb:linear}).  Thus, given $x_g$, once there gain matrix $K$ has been chosen, the vector $v$ must be found to satisfy the equation

\begin{equation}
(A+BK)x_g+Bv=0
\label{OffSet}
\end{equation}

\begin{proposition}
Let (\ref{eq:jb:linear}) be controllable, and let the $m\times n$ matrix $K$ be chosen such that the eigenvalues of $A+BK$ are in the open left half plane and the vector $v$ is is chosen to satisfy (\ref{OffSet}).  Then the $m$ control inputs
\begin{equation}
u_j(t)=v_j + k_{j1}x_1(t) + \cdots + k_{jn}x_n(t)
\label{eq:jb:standard}
\end{equation}
steer (\ref{eq:jb:linear}) toward the goal $x_g$.
\end{proposition}

We note that under the assumption that $m>n$, the values of both the matrix $K$ and the vector $v$ are underdetermined.  Hence there is some flexibility in parametric exploration of the design (\ref{eq:jb:standard}).  Assuming the matrix $B$ has full rank $n$, we can define the $m\times n$ matrix $R=B^T(BB^T)^{-1}A$.  Suppose $x_g\in\mathbb{R}^n$ is a desired goal to which we wish to steer the system (\ref{eq:jb:linear}).  Write
\begin{equation}
(R+K)x_g +\left(\begin{array}{c}
v_1\\
\vdots\\
v_m\end{array}
\right)= \left(\begin{array}{c}
0\\
\vdots\\
0\end{array}
\right),
\label{OffSet1}
\end{equation}
where
\[
K=\left(\begin{array}{ccc}
k_{11} & \dots & k_{1n}\\
\vdots & \dots & \vdots\\
k_{m1} & \dots & k_{mn}
\end{array}\right)
\]
is an $m\times n$ matrix of gain coefficients determined so as to place the poles of $A+BK$ at desired positions in the open left half plane. This determines the offset vector $v$ uniquely, but the resulting control inputs will not have the property of being robust with respect to channel intermittency.

To examine the effect of channel unavailability and channel intermittency, let $P$ be an $m\times m$ diagonal matrix whose diagonal entries are $k$ 1's and $m$-$k$ 0's.  For each of the $2^m$ such projection matrices, we shall be interested in cases where $(A,BP)$ is a controllable pair.  We have the following:

\begin{definition}\rm
Let $P$ be such a projection matrix with $k$ $1's$ on the main diagonal.  The system (\ref{eq:jb:linear}) is said to be $k$-{\em channel controllable with respect to} $P$ if for all $T>0$, the matrix
\[
W_P(0,T)= \int_0^T\, e^{A(T-s)}BPB^T{e^{A(T-s)}}^T\, ds.
\]
is nonsingular.
\label{def:jb:One}
\end{definition}

\smallskip




\begin{example} \rm
Consider again the three input system
\begin{equation}
\left(\begin{array}{c}
\dot x_1 \\
\dot x_2\end{array}\right)
=\left(\begin{array}{cc}
0 & 1\\
0 & 0\end{array}
\right)
\left(\begin{array}{c}
x_1 \\
x_2\end{array}\right)
+\left(
\begin{array}{ccc}
 0 & 1 & 1 \\
 1 & 0 & 1 \\
\end{array}
\right) \left(
\begin{array}{c}
 u_1 \\
 u_2 \\
 u_3 \\
\end{array}
\right).
\label{eq:jb:kChannel}
\end{equation}
Adopting the notation
\[
P[i,j,k]=
\left(
\begin{array}{ccc}
 i & 0 & 0 \\
 0 & j & 0 \\
 0 & 0 &k
\end{array}\right),
\]
the system (\ref{eq:jb:kChannel}) is 3-channel controllable with respect to $P[1,1,1]$; it is 2-channel controllable with respect to $p[1,1,0],P[1,0,1],$ and $P[0,1,1]$.  It is 1-channel controllable with respect to $P[1,0,0]$ and $P[0,0,1]$, but it fails to be 1-channel controllable with respect to $P[0,1,0]$.
\end{example}

It is natural to ask whether systems that are $k$-channel controllable are also $k$-{channel stabilizable}  in the sense that loss of certain channels will not affect either the goal equilibrium or its stability.  Once the $k_{ij}$'s are chosen as stabilizing feedback gains, the control offsets $v_i$ can be determined by solving (\ref{OffSet}).  With $m>n$, this is an underdetermined system, a particular solution of which is given by (\ref{OffSet1}).  To exploit the advantages of a large number of control input channels, we turn our attention to using the extra degrees of freedom in specifying the control offsets $v_1,\dots,v_m$ so as to make ({\ref{eq:jb:linear}) resilient in the face of channels being intermittently unavailable.
To see where this may be useful, consider using (\ref{OffSet1}) to steer (\ref{eq:jb:kChannel}) toward 
goal points on the unit circle.  We compare $(\cos\phi,\sin\phi)$ when $\phi=0$ and $\phi=\frac{\pi}2$.  If any two of the three input channels are operating, then the control inputs defined by these $k_{ij}$'s and $v_i$'s steer the system to the $\phi=0$ target (1,0).  But if any of the three channels is missing, these controls will fail to approach the target (0,1).  These cases are illustrated in Fig.\ \ref{fig:jb:2Channel}.

We note that the offset values $v_i$ that satisfy (\ref{OffSet1}) also satisfy
$B\left[(\hat A + K)x_g +v\right]=0$
where $\hat A$ is any $m\times n$ matrix satisfying $B\hat A = A$.  Under the assumption that $B$ has full rank $n$, such matrix solutions can be found--although such an $\hat A$ will not be unique.  Once $\hat A$ and the gain matrix $K$ has been chosen, the offset vector $v$ is determined by the equation 
\begin{equation}
(\hat A + K)x_g+v=0.
\label{OffSet2}
\end{equation}
The following gives conditions under which $\hat A$ may be chosen to make (\ref{eq:jb:linear}) resilient to channel dropouts.

\begin{theorem}
Consider the linear system (\ref{eq:jb:linear}) in which the number of control inputs, $m$, is strictly larger than the dimension of the state, $n$ and in which rank $B=n$.  Let the gain $K$ be chosen such that $A+BK$ is Hurwitz, and assume that
\begin{description}
\item {i\rm )} $P$ is a projection of the form considered in Definition \ref{def:jb:One} and (\ref{eq:jb:linear}) is ${\ell}$-channel controllable with respect to $P$;
\item{ii\rm )} $A+BPK$ is Hurwitz;
\item{iii\rm )} the solution $\hat A$ of $B\hat A=A$ is invariant under $P$---i.e., $P\hat A = \hat A$; and
\item{iv)\rm } $BP$ has rank $n$.
\end{description}
The the control inputs defined by (\ref{eq:jb:standard}) steer (\ref{eq:jb:linear}) toward the goal point $x_g$ whether or not the $(m-\ell)$ input channels that are mapped to zero by $P$ are available.
\label{thm:jb:four}
\end{theorem}

\noindent{\em Proof:} \ 
Under the stated assumptions, the point $x_g$ is a stable rest point of (\ref{eq:jb:linear}).  If $(m-\ell)$ channels associated with the projection $P$  are unavailable, the evolution of  (\ref{eq:jb:linear}) becomes
\[ 
\dot x=Ax+BPu = B(\hat Ax+Pu).
\]
Because $\hat A$ is invariant under $P$, this may be rewritten as
\[
\dot x=BP(\hat A x +u),
\]
and with $v$ defined by (\ref{OffSet2}), this can also be rendered as
\[
\dot x = (A+BPK)x + BPv.
\]
The goal $x_g$ is the unique stable rest point of this system.
\begin{flushright}$\Box$\end{flushright}

\medskip 

\begin{example} \rm
For (\ref{eq:jb:kChannel}) (the system considered in Example 2),
define control inputs of the form (\ref{eq:jb:standard}) where the gains are chosen to simultaneously stabilize the three two-input systems
\[
\left(
\begin{array}{c}
 \dot x_1\\
 \dot x_2 \end{array}
 \right)
  = 
\left(\begin{array}{cc}
  0 & 1\\
  0 & 0\end{array}
  \right) \left(
\begin{array}{c}
 x_1\\
 x_2 \end{array}
 \right) 
 + \left(\begin{array}{cc}
k_{21} & k_{22}\\
k_{11} & k_{12}
\end{array}
\right)\left(\begin{array}{c}
x_1\\ x_2 \end{array}\right),
\]

\[
\left(
\begin{array}{c}
 \dot x_1\\
 \dot x_2 \end{array}
 \right)
  = \left(\begin{array}{cc}
  0 & 1\\
  0 & 0\end{array}
  \right) \left(
\begin{array}{c}
 x_1\\
 x_2 \end{array}
 \right)
 +  \left(\begin{array}{cc}
k_{31} & k_{32}\\
k_{11} + k_{31} & k_{12} +k_{32}
\end{array}
\right)\left(\begin{array}{c}
x_1\\ x_2 \end{array}\right),
\]

and 

\[
\left(
\begin{array}{c}
 \dot x_1\\
 \dot x_2 \end{array}
 \right)
 =   \left(\begin{array}{cc}
  0 & 1\\
  0 & 0\end{array}
  \right) \left(
\begin{array}{c}
 x_1\\
 x_2 \end{array}
 \right)
  +  \left(\begin{array}{cc}
k_{21} + k_{31} & k_{22} + k_{32}\\
k_{31} & k_{32}
\end{array}
\right)\left(\begin{array}{c}
x_1\\ x_2 \end{array}\right).
\]
\smallskip

\noindent For any choice of LHP eigenvalues, this requires solving six equations in six unknowns, and choosing to place the eigenvalues in all cases at $s_1=-1,s_2=-1$ yields the values $k_{11}=0 ,k_{12}= -1,k_{21}= -1,k_{22}= 0, k_{31}= -1/2,k_{32}= -1/2$.  We note that for these choices of gain parameters and controls (\ref{eq:jb:standard}), that the full three-input closed loop system also has all poles in the open left half plane.  As in Example 1, the desired goal points will be on the unit circle: $(\cos\phi,\sin\phi)$.  Once the $k_{ij}$'s are determined, we can use (\ref{OffSet}) to solve for $v_1,v_2,v_3$ as functions of $\phi$.  A particular solution to (\ref{OffSet}) is given by (\ref{OffSet1}), and for this choice of stabilizing $k_{ij}$'s, the offsets are $(v_1,v_2,v_3) = (4/3\sin\phi,\cos\phi-2/3\sin\phi,1/2\cos\phi+1/6\sin\phi)$.  As noted above, these values do not generally steer (\ref{eq:jb:kChannel}) toward the goal in cases where an input channel is unavailable.

An alternative design approach utilizing Theorem \ref{thm:jb:four} is to rewrite (\ref{OffSet}) as 
\[
B\left[(\hat A + K)x_g+v\right]=0,
\]
where $\hat A$ is a solution of $B\hat A = A$.  For the system (\ref{eq:jb:kChannel}), these solutions constitute a two parameter family:
\[
\hat A= \left(\begin{array}{cc}
-s & -\frac13 - t\\
-s & \frac23 - t \\
s & \frac12 +t \end{array}
\right),
\]
with the values $s=t=0$ giving the particular solution $\hat A = R$ of equation (\ref{OffSet1}).  Checking for the invariance of $\hat A$ under the various projections $P[i,j,k]$, we find that $\hat A=\left(\begin{array}{cc}
0 & 0\\
0 & 1\\
0 & 0\end{array}
\right)$ is invaraiant under $P[1,1,1], \ P[0,1,1],$ and $P[1,1,0]$, but not under $P[1,0,1]$.  The controls defined by the given $k_{ij}$'s and satisfying (\ref{OffSet2}) for this choice of $\hat A$ have the offsets $(v_1,v_2,v_3)=(\sin\phi,\cos\phi-\sin\phi,\frac12\cos\phi+\frac12\sin\phi)$.  Using Theorem \ref{thm:jb:four} and as illustrated in Fig.\ \ref{fig:jb:2Channel}, the system (\ref{eq:jb:linear}) steered by the control inputs defined by (\ref{OffSet2}) reaches the desired goal state if either the first or the third channel drops out, but may fail to reach the target if the second channel drops out.
\end{example}

\begin{figure}[h]
\begin{center}
\includegraphics[scale=0.45]{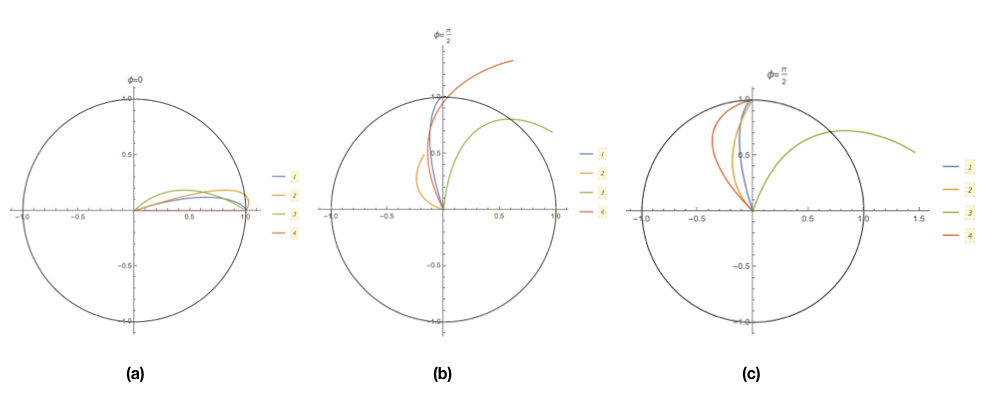}
\caption{Resilience to channel dropouts is illustrated.  In (a), the controls defined by (\ref{OffSet}) are seen to reach the goal point $(1,0)$ whether or not the channels for all three inputs are available.  In (b), however, the controls defined by (\ref{OffSet}) fail to reach the goal point (0,1) if any channel becomes unavailable.  For the goal point (1,0), both (\ref{OffSet}) and (\ref{OffSet2}) define the same controls, but for the goal point (0,1), the controls defined by (\ref{OffSet2}) are resilient to dropouts of either the first or third channel, but not to dropouts of the second.}
\end{center}
\label{fig:jb:2Channel}
\end{figure}
We conclude by asking whether a small number of standard inputs can be modulated to produce motion toward a goal that differs from the goal points toward which any single input steers the system.
For instance, if we consider the control inputs of the previous example that were designed to reach the goal points (1,0) and (0,1), we can examine switching between these with the objective of steering the system to the point $(\frac{1}{\sqrt 2},\frac{1}{\sqrt 2})$.  We have considered simple modulation in which the inputs are switched in and out according to Markovian switching schemes with a variety of probabilities of switching.  Generally speaking the results were not encouraging, but using Markovian switching between the second and third channels produced a trajectory that 
came within 0.002 units of the goal $(1/\sqrt{2},1/\sqrt{2})$.  Work on modulation of (possibly large) finite sets of standard inputs is ongoing. 


\section{Conclusions and further work}

The paper has illustrated robustness and functionality advantages of having large numbers of input and output channels as well as benefits in being able to select activation patterns of simple control actions chosen from a catalogue.  The examples have been low dimensional, but the qualitative features are expected to be present in higher dimensional systems.  Further research is needed, however, to understand general approaches to feedback control and observer synthesis based on learning activation patterns in high dimensions.  Given our projection operator formulation of the channel intermittency problem, we expect that subspace methods (\cite{Ho}) and  the types of learning that have  been successful in automated text recognition and visual search (\cite{Sivic}) will be of use going forward.  Much of the work has been motivated by a desire to understand how the control algorithms we have proposed for optical flow-based robot navigation degrade when visual features become sparse and when advisable to switch from, say, flow balancing along corridor walls to some other algorithm that depends of different visual cues.  The common thread is the goal of ``learning to move by moving to learn.''

\smallskip

{\sc Acknowledgment:} This work has benefitted enormously from conversations with J. Paul; Seebacher, Laura Corvese, and Shuai Wang.

\bibliography{references}
\bibliographystyle{IEEEtran}

\end{document}